%% file: main.tex
\documentclass[letterpaper, 10 pt, conference]{ieeeconf}  % Comment this line out if you need a4paper

\IEEEoverridecommandlockouts                              % This command is only needed if
\usepackage{algorithm}
\usepackage[noend]{algorithmic}
\usepackage{graphicx}
\usepackage[usenames,dvipsnames]{xcolor}
\usepackage[font=small]{caption}
\let\labelindent\relax
\usepackage{subfig, float, amsmath, amssymb, amsfonts, graphics, graphicx, enumitem}
\usepackage{listings,multicol}
\usepackage{lipsum}
\usepackage[utf8]{inputenc}
\usepackage{wrapfig}
\usepackage{hyperref}

\usepackage{dashrule}

\newtheorem{defn}{Definition}
\newtheorem{prop}{Proposition}

\newtheorem{thm}{Theorem}

\newcommand{\Pref}{\textbf{BPref}}
\newcommand{\Proj}{\textbf{Proj}}
\newcommand{\xx}{\textbf{x}}
\newcommand{\aaa}{\textbf{a}}
\newcommand{\dd}{\textbf{d}}
\newcommand{\ww}{\textbf{w}}
\newcommand{\qq}{\textbf{q}}

\title{\LARGE \bf
Correct-by-construction control synthesis for buck converters with event-triggered state measurement
}

\author{ Liren Yang, Xiaofan Cui, Al-Thaddeus Avestruz, Necmiye Ozay  % <-this % stops a space
\thanks{Authors are with the Dept. of
       EECS, Univ. of Michigan, Ann Arbor, MI 48109, USA
      {\tt \{yliren, cuixf, avestruz, necmiye\}\small @umich.edu}. This work is supported by NSF grants CNS-1446298 and ECCS-1553873, and by US Army CCDC Ground Vehicle Systems Center under
agreement W56HZV-14-2-0001. DISTRIBUTION A. Approved for public release; distribution unlimited. (OPSEC \#1802). This is an extended version of the paper \cite{yang2019correctacc} to appear in the Proceedings of the 2019 American Control Conference.
}
}

\begin{document}
\maketitle
\thispagestyle{empty}
\pagestyle{empty}

\begin{abstract} 
In this paper, we illustrate a new correct-by-construction switching controller for a power converter with event-triggered measurements. The event-triggered measurement scheme is beneficial for high frequency power converters because it requires relatively low-speed sampling hardware and is immune to unmodeled switching transients. While providing guarantees on the closed-loop system behavior is crucial in this application, off-the-shelf  abstraction-based techniques cannot be directly employed to synthesize a controller in this setting because controller cannot always get instantaneous access to the current state. As a result, the switching action has to be based on slightly out-of-date measurements. To tackle this challenge, we introduce the out-of-date measurement as an extra state variable and project out the inaccessible real state to construct a belief space abstraction. The properties preserved by this belief space abstraction are analyzed. Finally, an abstraction-based synthesis method is applied to this abstraction. We demonstrate the controller on a constant on-time buck voltage regulator plant with an event-triggered sampler. The simulation verifies the effectiveness of our controller.
\end{abstract}

%%%%%%%%%%%%%%%%%%%%%%%%%%%%%%%%%%%%%%%%%%%%%%%%%%%%%%%%%%%%%%%%%%%%%%%%%%%%%%%%
\section{Introduction}\label{sec:Intro}
\input{sec_Intro.tex}

\section{System Model, Requirements and Problem Statement}\label{sec:ProbState}
\input{sec_ProbState.tex}

\section{Solution approach}\label{sec:Sol}
\input{sec_SolApproach.tex}

\section{Simulation Results}
\input{sec_Result.tex}

\section{Conclusions} % and Future Work}
The contributions of this paper are twofold. First, we developed a hybrid automaton model for constant-on time buck converter that integrates the circuit model with physical requirements. Second, we proposed a belief space abstraction-based control synthesis scheme for the buck converter with event-triggered measurements that guarantees satisfaction of functional requirements. The correctness of the control synthesis scheme is analyzed and the resulting controller is shown to outperform a baseline design on a simulation example. Ideas exploiting the problem structure to reduce complexity and conservatism are presented, which we believe can be relevant in other application domains as well.

Although the abstraction (hence the controller) size is modest compared to the abstractions in other real applications, the practical implementation still requires to further simplify the control law. We will consider extracting a simplified switching rule from the obtained correct-by-construction controller and testing it on hardware as part of our future work. 

%{\small
%\noindent{\bf Acknowledgments:}
%% We thank our colleague Xiaofan Cui from the Dept. of EECS, Univ. of Michigan, who provided insight and expertise  that greatly helped us understand the power electronics background, and assisted in preparing the parts in this draft that are related to the buck converter. 
%This work is supported by NSF grants CNS-1446298 and ECCS-1553873, and by U.S. Army TARDEC under
%agreement W56HZV-14-2-0001.}

\bibliographystyle{abbrv}
\bibliography{main}

\section*{Appendix}\label{sec:app}
\input{sec_Appendix.tex}

\end{document}

%% file: sec_Intro.tex
%!TEX root = main.tex

% sec_Intro
{\color{black}
There is an increasing demand for higher performance power converters; digital controllers with the promise of higher speed and lower price thanks to Moore's law look to fulfill this challenge. Compared to its analog counterpart, a digital controller is more flexible in accommodating different high performance applications. Besides, more complicated control algorithms can be more easily realized by digital \cite{peterchev2003architecture}.

Traditional digital controllers for power converters require a fixed-rate digital sampler for measurements, with a sampling rate much higher than the actuation frequency.\footnote{In power converters, actuation is typically via controlling switches in the system; and sampler (or sampling hardware) refers to the measurement mechanisms.} This is because the digital control algorithms employed usually require high-fidelity estimates of system states.  Hence, with the increasing demand on dynamic response and switching frequency of power converters, sampling hardware complexity increases dramatically \cite{burd2000dynamic}. In addition, for variable switching-frequency power converters, sampling measurements at a fixed-rate is susceptible to distortion in measurement quality due to unmodeled switching transients. As an alternative, event-driven control is starting to draw attention in power electronics as it can alleviate the requirement for high complexity sampling hardware without sacrificing the performance significantly \cite{cui2018anewframe,rathore2016event}. 

%Traditional digital controllers for power converters are founded upon a fixed rate digital sampler with a sampling rate much higher than the actuation frequency (i.e.,  the switching frequency\footnote{Many power converters are switched systems by their nature. and actuation hence means switching.}), because the digital control algorithm is usually based on high fidelity reconstruction of state trajectories. Hence, with the increasing demand on dynamic response and switching frequency of power converters, }{\color{red}sampling }{\color{blue} hardware complexity increases dramatically \cite{burd2000dynamic}. For variable switching-frequency power converters, another problem is that the fixed-rate sampling action might coincide with the switching action and measurement can be disturbed by unmodelled switching transient.

%Event-driven control in power electronics is starting to draw people's attention. \cite{cui2018anewframe} exhibits a buck voltage regulator using a switching-synchronized sampler with a linear discrete controller. The converter shows fast dynamic response with low requirement on }{\color{red}sampling }{\color{blue} hardware complexity. \cite{rathore2016event} shows an event-driven controller in power converters,  which reduces unnecessary switchings at the cost of only a small degradation in performance, while improving the converter efficiency. 

In this work, we consider controller design for a specific power converter called constant on-time buck converter.
Such converters are widely used in voltage regulator modules and point-of-load converter applications \cite{bari2017enhanced}. 
The converter usually has two switches as actuators, denoted as $S_1$ and $S_2$ respectively, and the term ``constant on-time" indicates that the turn-on time of the switch $S_1$ is kept constant. Constant on-time is shown to provide robustness to the circuit parameter variations \cite{galvez2011improvements} and avoid chattering in hysteresis control \cite{banerjee2001nonlinear}.
%In this mode, the steady state frequency is only determined by the constant on-time, input voltage and output voltage, and hence is more robust to the circuit parameter variation comparing to sliding mode control \cite{galvez2011improvements}. Moreover, the problem of infinite steady state switching frequency in hysteresis control does not occur here \cite{banerjee2001nonlinear}.
A notable feature of the constant on-time buck converter considered in this work is that its states are measured in an event-triggered manner. By imposing certain physical constraints on when the voltages and currents are measured, the burden on sampling hardware can largely be reduced. However, this non-uniform sampling mechanism necessitates new control algorithms that can work with limited and possibly out-of-date measurements. 
%Since the off-time is controlled in every switching cycle, the switching frequency is not constant, especially at the transient phase.  
%Here, we keep our measurement synchronous to this varying switching frequency. Specifically, the measurement of the output capacitor voltage is kept in the middle of the on-time, and the inductor current is measured during off-time. This switching-synchronized measurement is that it largely release the burden on digital hardware because the output voltage sampling rate as low as 1 sample per switching cycle. Moreover, our measurement can naturally avoid the switching transient.

%The controller together with a constant on-time buck VRM is verified through simulation and results show that controller can effectively regulate the output voltage in a tight bound.

}

% intro to c-b-c
In order to address this challenge, we formulate the constant on-time buck converter control problem as a temporal logic game of switched affine systems and solve the game for an abstraction of the system \cite{beltabook,nilsson2017augmented}. The solution of the game leads to a correct-by-construction switching controller that assures a desired closed-loop behavior specified using temporal logics \cite{baier2008principles}, making it possible to strictly regulate the systems' behavior. Moreover, this set of approaches is particularly good at handling switched systems like the buck converter considered in this paper. However, standard approaches for abstraction-based synthesis usually assume a uniform sampling rate and immediate access to the system states, assumptions that do not hold for the buck converter system. Controller design for systems like this can be handled under the framework of sampled-data systems, with extra analysis to guarantee the time robustness, e.g., \cite{khammash1993necessary}. 
In particular,  abstraction-based synthesis of such system with incremental stability assumption are studied under the framework of sampled-data systems in recent works like \cite{kader2018symbolic}, where one can tolerate nonuniform but bounded sampling rates, while still enjoying the approximate-bisimulation property thank to the incremental stability of the system.

The main difference of our approach from \cite{kader2018symbolic} is that we solve the problem as a temporal logic game with imperfect (particularly, delayed) information \cite{chatterjee2006algorithms,dimitrova2008abstraction}. With this approach, we do not need to assume an incrementally stable system or finite sampling period (the case with infinitely long sampling period may be rare in practice but there can be potential situations where further measurement is not helpful for the decision making). The main challenge with the imperfect information game solving is that it usually involves the construction of a belief space that grows exponentially \cite{chatterjee2006algorithms}. 
% Works like \cite{mickelin2014synthesis} try to avoid such belief space construction using bounded  error estimation techniques. 
In this paper, we provide belief space reduction schemes tailored to the specific features in the buck converter problem.

%% file: sec_ProbState.tex
%!TEX root = main.tex

In this section, we define the control problem by giving the buck converter model and the requirements of the system.
%In this paper, we consider two categories of requirements: physical requirements and functional requirements. 
We consider two types of requirements: physical requirements and functional requirements.
Physical requirements refer to the extra restrictions on the measurement mechanism in order to
improve the system performance and reduce the hardware implementation cost, as discussed in the introduction, whereas
functional requirements specify the desired closed-loop system behavior.
%While function requirements specifies the desired closed-loop system behavior, physical requirements refer to the extra restrictions on the controller structure that are related to improving the system performance and reducing the hardware implementation cost, as discussed in the introduction. 
%In particular, the event-triggered measurement scheme of the buck converter system is resulted from these physical requirements. 

% Model
\subsection{Physical Requirements and Buck Converter Model}\label{sec:sys_model}
% Naive cont-time model 
%The buck converter system without extra physical requirements is modeled as a switched affine system. 
%Then we present an hybrid automaton that integrates the physical requirements with this model. 
We first develop a hybrid automaton model that integrates the buck converter circuit model with
the measurement scheme derived from the physical requirements. While hybrid automata are
used to model power converters for reachability-based  verification purposes in the past \cite{beg2017model,kuhne2010analysis}, to 
the best of our knowledge, no prior work exists that considers event-triggered measurements.
%Using hybrid automaton to model power converters are not new and enables formal verification based on reachable set computation, see,  for example, \cite{beg2017model}, \cite{kuhne2010analysis}. However, no works were found to consider event-triggered measurement scheme, which is the main focus of this paper. 

\begin{figure}[h]
  \centering
  \includegraphics[width=3.3in]{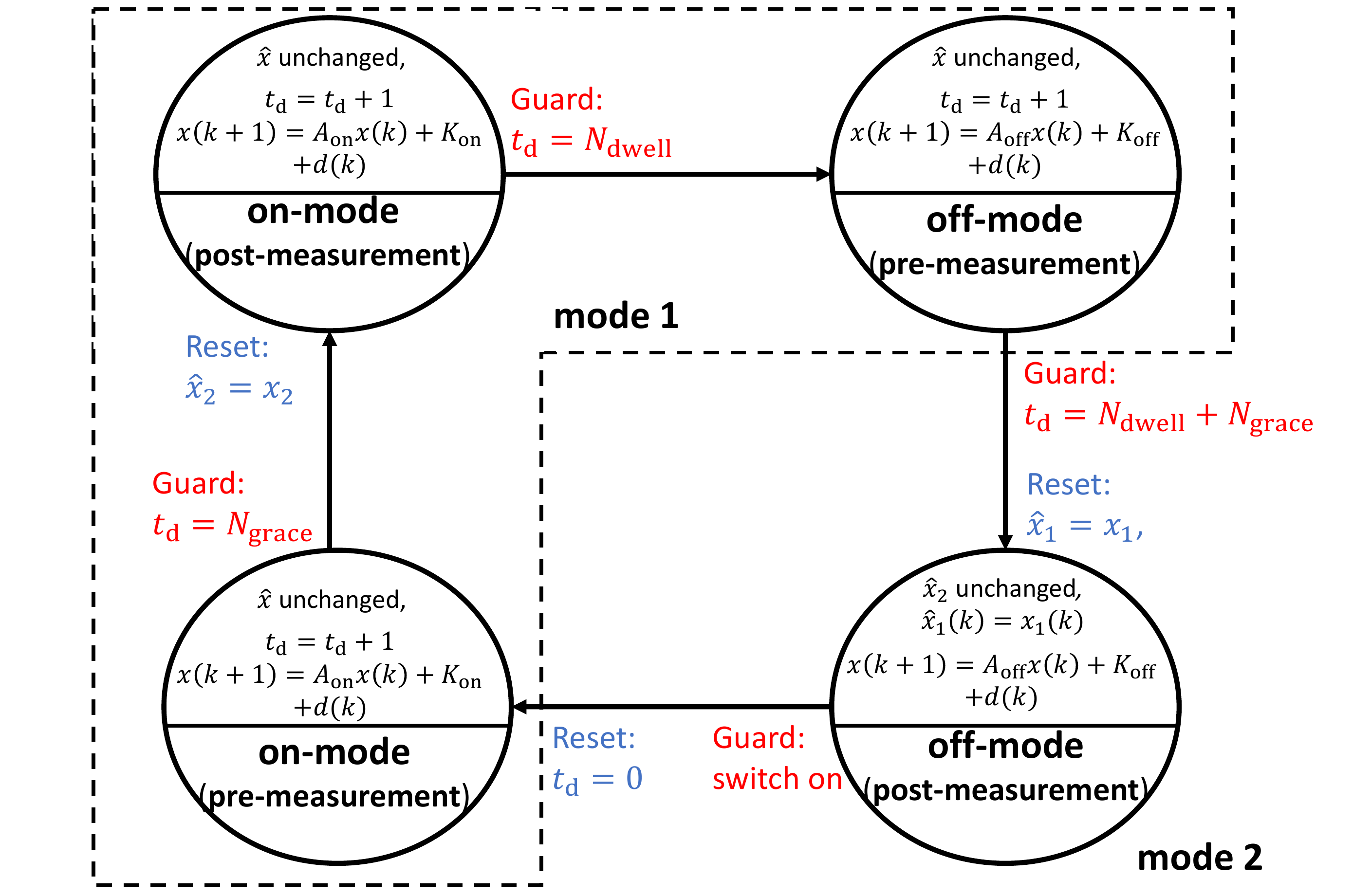}\\
  \caption{Hybrid automaton $\mathcal{H}$ that models the buck converter with the switching and measurement restrictions.}\label{fig:HAuto}
  \vspace{-2mm}
\end{figure}

{\color{black} The buck converter has two switches as actuators, denoted as $S_1$ and $S_2$ respectively.
We describe the converter's model by a switched affine system with two modes, and we refer the two modes as the ``on-mode'' and the ``off-mode''.
The ``on-mode'' corresponds to the case when $S_1$ is on and $S_2$ is off, while the ``off-mode'' corresponds to the case when $S_1$ is off and $S_2$ is on\footnote{The both-on combination leads to a voltage source being short circuited and the both-off combination leads to an open circuit current source. Hence these two configurations are never used.}. 
The continuous-time circuit dynamics under the on-mode is given by
\begin{align}
\tfrac{d}{dt}\left[
\begin{array}{c}
I \\
V \\
\end{array}
\right]
 =  \left[
\begin{array}{cc}
 0 & -\tfrac{1}{L} \\
 \tfrac{1}{C} & -\tfrac{1}{RC}\\
\end{array}
\right] 
\left[
\begin{array}{c}
I \\
V \\
\end{array}
\right]
 +  
\left[
\begin{array}{c}
\tfrac{V_{\rm in}}{L} \\
0 \\
\end{array}
\right], 
\label{eq:Dyn_cont}
\end{align}
where the variables and parameters are given in TABLE \ref{tab:var}. 
\begin{table}[h] 
	\centering 
	\caption{Parameter \& Variables }  \label{tab:var}
	\begin{tabular}{c c c}
		\textbf{Notation} & \textbf{Physical meaning} & \textbf{Value} \\ \hline
		$I$ &  inductor current  & - \\
             $V$ &  capacitor voltage  & - \\
\hline
        $C$  &  output capacitor  & $2.5\times 10^{-3}$ (F) \\ 
        $L$  &  output inductor  & $2\times 10^{-4}$ (H) \\ 
        $R$ &  load resistor   & 0.5 ($\Omega$) \\ 
        $V_{\rm in}$ & input voltage &  100 (V) \\ 
\hline
	\end{tabular}
\end{table} 
The dynamics under the off-mode can be defined by replacing the rightmost constant offset term in Eq. \eqref{eq:Dyn_cont} by $[0,0]^T$. 
%In the rest part of the paper, we will consider the the system sampled with time period $\Delta t = 25$ns. 
With a sampling time of $250$ ns, the discrete-time switched dynamics are given by: 
\begin{align}
x(k+1) = A_{s} x(k) + K_{s} + d
\end{align}
where $x = [x_1, x_2] = [I, V]\in  \mathbb{R}^2$ is the state, $s \in \{\text{on, off}\}$ is the mode, $d \in D\subseteq \mathbb{R}^2$ is the bounded disturbance. 

% Restrictions on mode switching & state measurement
As alluded to in the introduction, for some performance and cost considerations, %the system has further restriction on mode switching for some performance considerations, and its state measurement is related to the switching signal and hence event-triggered.  
the following physical requirements are imposed on the mode switching and state measurements: 
\begin{itemize}[nolistsep]
\item The dwell time of the on-mode is fixed as $T_{\rm dwell}$ as per constant-on time converter scheme. For the specific voltage regulator in this paper, we set $T_{\rm dwell} = 12 \ \mu \text{s}$  (i.e., $N_{\rm dwell} = 48$  samples) for illustration. 

\item The state $V$ is measured only once per cycle after the system has stayed in the on-mode for exactly $T_{\rm grace}$. 
%The exact $T_{\rm grace}$ allows the measurement instant to be fully synchronized (i.e. to occur with a same frequency and a fixed phase difference) to the switching-on events. 
Although $V$ can be accessed by more than once per switching cycle, voltage measurement is very costly \cite{ltc2378-16}. %For example, a high-end commercial 16-bits ADC can finish 1 voltage measurement in 75 ns \cite{ltc2378-16}. A 5 MHz power converter has a switching period is 200 ns and only 2 voltage measurement can be finished in 1 switching cycle.
Therefore, one of the aims of this paper is to do as few voltage measurements as possible. 

\item The dwell time of the off-mode is lower bounded by $T_{\rm grace}$ for allowing states to settle from the switching transient. %For the specific voltage regulator in this paper, 
We set $T_{\rm grace} = 1 \ \mu\text{s}$ (i.e., $N_{\rm grace} = 4$ samples) in this paper.
The state $I$ can be measured at any time after $T_{\rm grace}$ during off-mode. It is much easier to measure the current during the off-mode rather than the on-mode because during on-time, the inductor current has to be measured by a differential voltage sensor which is harder to implement.
%requires much efforts to implement. 
One difference from voltage measurements is that the current measurement can be done for more than once during off-time because the cost of current measurement is not as high as voltage measurement. %The reason is for a voltage regulator application, a high resolution voltage sensor is required to satisfy the tight output voltage regulation bound. Any noises or uncertainty on the voltage sensor is a reference disturbance which controller cannot reject. However, the current sensor can be relatively low resolution because the controller will automatically correct the disturbance caused by uncertain measurement and set the current value to the appropriate position. 

 \item After staying in the on-mode for $T_{\rm dwell}$, the system is forced to switch to off-mode so that the inductor current information can be updated during every cycle.
\end{itemize}}
To capture the dynamics of the buck converter system with the above switching and measurement restrictions, we introduce two extra variables: 1)  $\hat{x} = [\hat{x}_1, \hat{x}_2]$ that represents the measurement of the actual state $x = [x_1, x_2]$ , and 2) $t_{\rm d}$ that represents the time spent in the on- or off-mode.
With the extra variables, the model can be represeted with the hybrid automaton, denoted by $\mathcal{H}$, shown in Fig. \ref{fig:HAuto}. 
Note that the discrete state of hybrid automaton $\mathcal{H}$ can be further simplified, by combining the three discrete states in the dashed line box. We do this because all the transitions within the dashed line box are autonomous, while the only non-autonomous transition is between the states marked as ``off-mode (post-measurement)'' and ``on-mode (pre-measurement)''. 
From now on, we will refer to the combined state (i.e., the three discrete states in dashed line box) as mode 1, and the other discrete state as mode 2. 
To distinguish these two modes from the on/off-mode of the buck converter, we use $a\in A = \{1,2\}$ to denote the above simplified modes, and $s  \in \{\text{on, off}\}$ for the on/off-mode. 
We say a mode sequence $\textbf{a} = a(1)a(2)a(3)\dots$ is admissible for the hybrid automaton $\mathcal{H}$ if it evolves according to the following rules:  
\begin{itemize}[nolistsep]
\item $a(k) = 2 \Rightarrow a(k+1) = 1\text{ or } 2$. This means when $a(k) = 2$, one can either continue with $a(k+1) = 2$ or switch to $a(k+1) = 1$. 
%\item For any \textit{longest} fragment $a(k)\dots a(K)$ consists of only 1's, % such that $a(l)=1$ for $l = k,\dots, K$, 
%$(K-k+1)\text{ mod }(N_{\rm dwell} + N_{\rm grace}) = 0$ must hold. 
%By term ``longest'' we mean one of the following three conditions holds: (i) $a(k-1) = a(K+1) = 2$, or (ii) $k=1$ and $a(K+1)=2$, or (iii) $a(k-1)=2$ and $K = \infty$. 
%This means that once the mode is set to be 1, it will maintain there for $N_{\rm dwell}+N_{\rm grace}$ steps. 
%Note that the system will then automatically set to mode 2 according to Fig. \ref{fig:HAuto}, but we assume that the system can switch to mode 1 immediately after that. Hence there can be multiple consecutive fragments, each containing $N_{\rm dwell}+N_{\rm grace}$ 1's.  
\item {\color{black} For any \textit{maximal} 1-fragment $a(k)\dots a(K)$ consisting of only consecutive 1's, % such that $a(l)=1$ for $l = k,\dots, K$, 
$(K-k+1)\text{ mod }(N_{\rm dwell} + N_{\rm grace}) = 0$ must hold. 
The term ``maximal'' means that one of the following three conditions holds: (i) $a(k-1) = a(K+1) = 2$, or (ii) $k=1$ and $a(K+1)=2$, or (iii) $a(k-1)=2$ and $K = \infty$. 
This means that once the mode is set to 1, it will remain there for $N_{\rm dwell}+N_{\rm grace}$ steps. 
Note that the system will then automatically switch to mode 2 according to Fig. \ref{fig:HAuto}, but we allow controller to decide to skip ``off-mode (post-measurement)'' state with the reset $\hat{x}_1 = x_1$. Hence there can be multiple consecutive 1-fragments, each containing $N_{\rm dwell}+N_{\rm grace}$ terms. }
\end{itemize}
Under an admissible mode sequence $\textbf{a}$, the converter state $x$ and its measurement $\hat{x}$ update  accordingly: 
\begin{itemize}[nolistsep]
\item if $a(k) = 2$, the converter state $x$ will update with the off-mode dynamics; the measurement $\hat{x}_1$ updates at every step while $\hat{x}_2$ is always fixed;
\item if $a(k) = 1$, the converter state will evolve according to the on-mode dynamics whenever $t_{\rm d}\leq N_{\rm dwell}$, and then evolves with the off-mode dynamics whenever $t_{\rm d} > N_{\rm dwell}$;  the measurement $\hat{x}_1$ is always fixed, and $\hat{x}_2$ gets updated only when $t_{\rm d} = T_{\rm grace}$. 
\end{itemize}
We define $\textbf{x}^{\textbf{a},\textbf{d}} = \big(x(1),\hat{x}(1)\big)\big(x(2),\hat{x}(2)\big)\dots$ to be the unique execution of hybrid automaton $\mathcal{H}$ under action sequence $\textbf{a}$ and disturbance profile $\textbf{d} = d(1)d(2)d(3)\dots$. 
Note that $t_{\rm d}$ is not included as a state in the execution because it will be redundant if $\textbf{a}$ is given. However, we will need $t_{\rm d}$ when control action $a$ is to be determined. 
In the reset of the paper, we call state feedback\footnote{At this point, it does not make much sense to assume both $x$ and $\hat{x}$ to be accessible to the controller. However, as will be seen later in the paper, such controller is useful for stating the relation between the closed-loop hybrid automaton and its raw abstraction. }  controller $C:X\times \hat{X}\times \{0,\dots,N_{\rm dwell}+N_{\rm grace}\}\rightarrow 2^A$ to be admissible by $\mathcal{H}$ if $C(x,\hat{x},t_{\rm d}) = 1$ for $t_{\rm d} < N_{\rm dwell}+N_{\rm grace}$, in which case the system stays at mode 1 according to Fig. \ref{fig:HAuto}.

\subsection{Functional Requirements}
We use Linear Temporal Logic (LTL) to specify the functional requirements of the buck converter system, i.e., the desired closed-loop system behavior of the hybrid system $\mathcal{H}$. In what follows we briefly introduce the syntax and the semantics of LTL, and refer the reader to \cite{baier2008principles} for more details.

% syntax
\subsubsection{LTL Syntax}
Let $AP$ be a set of atomic propositions, the syntax of LTL formulas over $AP$ is given by
\begin{align}
\varphi :: = \pi \mid \neg\varphi \mid \varphi_1\vee\varphi_2 \mid \bigcirc \varphi \mid \varphi_1\ \mathcal{U} \ \varphi_2
\label{eq:LTLGrammar}
\end{align}
where $\pi\in AP$. With the grammar given in Eq. \eqref{eq:LTLGrammar}, we define the other propositional and temporal logic operators as follows: $\varphi_1\wedge \varphi_2 \doteq \neg(\neg \varphi_1 \vee \neg \varphi_2)$, $\varphi_1 \rightarrow \varphi_2\doteq \neg \varphi_1 \vee \varphi_2$, $\lozenge \varphi \doteq True \ \mathcal{U} \ \varphi$, $\Box \varphi \doteq \neg \lozenge \neg \varphi$, $\varphi \ \mathcal{W} \ \psi\doteq (\varphi \ \mathcal{U} \ \psi) \vee (\Box \neg \psi)$. 

\subsubsection{LTL Semantics}
% semantics $\xxx \models \varphi$
We define a word $\textbf{w} = w(1)w(2)w(3)\dots$ to be an infinite sequence of subsets of atomic propositions (i.e.,  $w(k) \subseteq AP$), and interpret an LTL formula over the set of words as follows: 
\begin{itemize}[nolistsep]
  \item $ \textbf{w} \models \pi$ if and only if (iff) $\pi \in w(1)$,
  \item $ \textbf{w} \models \neg \varphi $ iff $\textbf{w} \nvDash\varphi$,
  \item $ \textbf{w} \models \varphi_1\vee \varphi_2$ iff $\textbf{w} \models\varphi_1$ or $\textbf{w} \models\varphi_2$,
  \item $ \textbf{w} \models \bigcirc \varphi$ iff $\textbf{w}_2 \models\varphi$,
  \item $ \textbf{w} \models \varphi_1 \ \mathcal{U} \ \varphi_2$ iff $\exists s\geq 0: \textbf{w}_s \models \varphi_2$ and $\forall t < s: \textbf{w}_t \models \varphi_1$,
\end{itemize}
where $\textbf{w}_k = w(k)w(k+1)w(k+2)\dots$ is the suffix of word $\textbf{w}$ starting from its $k^{\rm th}$ element. 
Given an infinite word $\textbf{w}$ and an LTL formula $\varphi$, we say $\varphi$ holds for $\textbf{w}$ (or $\textbf{w}$ satisfies $\varphi$) iff $\textbf{w} \models \varphi$.

\subsubsection{Buck Converter Specification in LTL}
The objective of the controller is to steer the buck converter state $x$ into a given  target region $X_{\rm target}$, and let it stay in $X_{\rm target}$ forever once arrived. 
In the meantime, the state should never go outside the given domain $X$. 
Let $AP = \{\texttt{safe}, \texttt{target}\}$, the above specification can be written as the following LTL formula: 
{\small
\begin{align}
\Phi = (\Box \texttt{safe}) \wedge (\neg \texttt{target} \ \mathcal{U} \ \Box \texttt{target}). 
\label{eq:spec}
\end{align}
}% 
Let $\textbf{a}$ be a mode sequence that is admissible for hybrid automaton $\mathcal{H}$ and $\textbf{d}$ be the disturbance profile, and let $\textbf{x}^{\textbf{a},\textbf{d}} = \big(x(1),\hat{x}(1)\big)\big(x(2),\hat{x}(2)\big)\dots$ be the associated execution,  we say that execution $\textbf{x}^{\textbf{a},\textbf{d}}$ satisfies LTL formula $\Phi$ if $\textbf{w}(\textbf{x}^{\textbf{a},\textbf{d}}) \doteq \ell\big(x(1)\big)\ell\big(x(2)\big)\dots \models \Phi$, 
where $\ell: X\rightarrow 2^{AP}$ is an observation map such that $\texttt{safe} \in \ell(x) \Leftrightarrow x\in X$ and $\texttt{target}\in \ell(x)\Leftrightarrow X_{\rm target}$. 
Note that the labeling is only with respect to the true state $x$ and we are not interested in labeling the measurement state $\hat{x}$. 
% Similar as before, we use $\textbf{w}^C_{\mathcal{H}}$ to denote a word generated by execution $\textbf{x}^C$ under controller $C$. 

%{\color{red} to add: some discussion on why not eventually always target. may be useful as our system relation formal statement is very specific to reach-avoid-stay. }

\subsection{Problem Statement}
Given the hybrid automaton $\mathcal{H}$, the LTL specification $\Phi$ defined in Eq. \eqref{eq:spec}, and an initial state $x_{\rm init}$ (or a set of initial states $X_{\rm init}$)\footnote{We assume that the actual state $x$ is accessible at the initial time. This assumption is reasonable as there will be no unmodeled switching dynamics that spoil the measurement at the very beginning. As a result, the measurement state $\hat{x}$ of hybrid automaton $\mathcal{H}$ is such that $\hat{x} = x$ at the initial time.}, 
the goal is to find a switching controller under which all the executions of $\mathcal{H}$ starting from $X_{\rm init}$ and  $t_{\rm d} = N_{\rm dwell} + N_{\rm grace}$ (i.e., starting from mode 2) satisfy LTL formula $\Phi$. 
In particular, the control action needs to be determined based on the history of $\hat{x}$, but not $x$.
% Requirement

%% file: sec_SolApproach.tex
%!TEX root = main.tex

% general
To synthesize a correct-by-construction controller for the buck converter with the described event-triggered measurement mechanism, we use abstraction-based synthesis. We first construct a finite transition system that over approximates the behavior of the buck converter and the measurement state $\hat{x}_2$, then we project out state $x_2$ that is not directly accessible and this leads to an abstraction in the belief space.
%({\color{red} should I call the FTS itself a ``belief space'' or should I can it ``the abstraction in the belief space''?})
The control problem is then solved as a reach-avoid-stay game on the belief space abstraction. Finally, several techniques are presented to reduce the belief space abstraction size. %While we exploit the problem structure in the buck converter application for some of the steps, the presented ideas can be useful in other similar applications as well.
%Some of these techniques are specific to the buck converter application. 

% Abstraction
\subsection{Raw Abstraction}
%In this work, we will use finite transition systems as abstractions for the buck converter system.  
The buck converter system will be abstracted by a finite transition system $TS$, a  five-tuple $(Q, A, AP, \lambda, \tau)$, where $Q$ is a finite set of states, $A$ is a finite set of actions, $AP$ is a set of atomic propositions, $\lambda: Q\rightarrow 2^{AP}$ is the labeling function and $\tau: Q\times A \rightarrow 2^Q$ is the nondeterministic transition map. With a slight abuse of notation, we define $\tau(Q_1,A_1) = \bigcup_{q\in Q_1, a\in A_1}\tau(q,a)$ for $Q_1\subseteq Q$ and $A_1\subseteq A$. Given an action sequence $\textbf{a} = a(1)a(2)a(3)\dots$, an execution of the transition system is a sequence $ \textbf{q}^{\textbf{a}} = q(1)q(2)q(3)\dots$ such that $q(k+1)\in \tau\big(q(k),a(k)\big)$, and the word associated with this execution is $\textbf{w}(\textbf{q}^{\textbf{a}}) = \lambda\big(q(1)\big) \lambda\big(q(2)\big) \lambda\big(q(3)\big) \dots$.

% Given a controller $\overline{C}:Q\rightarrow 2^A$, we define a resulted execution and its associated word by $\textbf{q}^{\overline{C}}= \textbf{q}^{\textbf{a}} $ and  $\textbf{w}^{\overline{C}}_{TS} = \textbf{w}^{\textbf{a}}_{TS}$, where  $\textbf{a} = a(1)a(2)a(3)\dots$ is such that $a(k) = \overline{C}\big(q(k-1)\big)$. 

We first compute an abstraction of the system with the extended state $[x, \hat{x}]^T \in X \times \hat{X}$, where $\hat{X}$ is the domain of $\hat{x}$, the out-of-date measurement of $x$. 
%For a reason that will be discussed in the later parts, we call this abstraction the ``raw abstraction''. 
This abstraction is called the ``raw abstraction''.
The raw abstraction should over-approximate the behavior of the hybrid automaton $\mathcal{H}$ presented in Section \ref{sec:sys_model} so that if a controller is synthesized for the raw abstraction, it can be applied to the hybrid automaton and the correctness is preserved. However, nondeterministic abstractions based on one-step reachability relations typically lead to
spurious behaviors (especially, in the absence of bisimulation relations) since the nondeterminism accumulates \cite{beltabook,nilsson2017augmented}.
%To enjoy such behavior over-approximation, one usually needs the abstraction to simulate the hybrid automaton  under each control action, which are usually achieved by one-step reachable set computation in abstraction construction \cite{beltabook,nilsson2017augmented}
%However, an abstraction obtained in this way turned out to be too conservative, i.e., containing too many spurious behaviors that do not exist in the hybrid automaton, for this specific application problem. 
%Besides, if one-step reachable set computation is used to construct the abstraction, extra states are needed to track the dwell time during mode 1, which will increase size of the abstraction state space. 
Our \emph{key insight} to mitigate this problem for the buck converter system is to leverage the constant duration of mode 1 and do long-term reachable set computation to avoid spurious transitions. This also eliminates the need for extra states to track the dwell time in mode 1. The constructed abstraction is in a sense ``multi-rate" where different transitions correspond to different durations on the actual system. This requires extra care in labeling the abstraction in general, for which we exploit the special structure in specification $\Phi$ to simplify the process.
% and the unnecessary states for dwell time tracking, and ii) use the special structure in specification $\Phi$ to simplify labeling of abstraction. 
These constructions lead to a raw abstraction that approximates the behaviors of $\mathcal{H}$ in a weaker sense but is enough to preserve the correctness as long as achieving specification $\Phi$ is considered. In what follows, we denote this raw abstraction by $TS^{\rm R} = (Q, A, AP, \lambda, \tau)$ and define each component in the five-tuple according to the idea mentioned above. Then we will formally state the system relation between the hybrid automaton $\mathcal{H}$ and this raw abstraction $TS^{\rm R}$. 

\subsubsection{Raw Abstraction Construction} 
First, the action set  $A = \{1,2\}$ is the set of the simplified modes defined at the end of Section \ref{sec:sys_model}. The atomic proposition set $AP = \{\texttt{safe}, \texttt{target}\}$. 

Secondly, to define the state set $Q$, we partition the true two-dimensional state domain $X$ and the two-dimensional measurement  domain $\hat{X}$ in the same way using a grid. Each point in the grid is denoted by $x^i$ (or $\hat{x}^i$, respectively) when it is used to abstract $X$ (or $\hat{X}$, respectively), and corresponds to a closed rectangular region $R^{x^i}$ (or $R^{\hat{x}_j}$, respectively). 
Now, we can express the discrete state set $Q$ by $\big(\{x^1, x^2,\dots, x^N\}\times \{\hat{x}^1, \hat{x}^2,\dots, \hat{x}^N\}\big) \cup \{q_{\rm out}\}$, where 
$q_{\rm out}$ is an extra state that represents the region outside the domain. 
Now that each discrete state $q \in Q\setminus \{q_{\rm out}\}$ corresponds to a tuple $(x^i, \hat{x}^j)$, it naturally corresponds to a four-dimensional rectangular region $R_q = R^{x^i} \times R^{\hat{x}^j}$ in the continuous state space $X\times \hat{X}$.  
We also use $R_q^x$ ($R_q^{\hat{x}}$ respectively) to denote the projection of $R_q$ onto $X$ ($\hat{X}$ respectively) space, that is, $R_q^x = R^{x^i}$ and $R_q^{\hat{x}} = R^{\hat{x}^j}$ for $q = (x^i, \hat{x}^j)$. 
Similarly, $R_q^{x_1}$, $R_q^{x_2}$, $R_q^{\hat{x}_1}$, $R_q^{\hat{x}_2}$ correspond to the one-dimensional projections of the four-dimensional rectangle $R_q$. 
%To avoid even heavier notations, we use Fig. \ref{fig:grid} to illustrate a two-dimensional grid partition and the related concepts explained above, and the idea should be clear for higher dimensional cases. 
We define the labeling function $\lambda$ to be such that $\texttt{safe} \in \lambda(q) \Leftrightarrow q\neq q_{\rm out}$, and $\texttt{target}\in \lambda(q) \Leftrightarrow R_q^x \subseteq X_{\rm target}$. 
In particular, we assume the grid partition of $X$ to be proposition preserving, i.e., for all $q\in Q\setminus\{q_{\rm out}\}$, either $\text{int}(R_q^x)\cap X_{\rm target} = \emptyset$ or $\text{int}(R_q^x)\subseteq X_{\rm target}$, where $\text{int}(R_q^x)$ is the interior of rectangle $R_q^x$.  
We also define target state set $Q_{\rm target} \doteq \{q\in Q\mid \texttt{target} \in \lambda(q)\}$ and the  initial state set $Q_{\rm init} \doteq \{q\in Q\mid R_q^{x} \cap X_{\rm init} \neq \emptyset\}$. 
%\begin{figure}[h]
%  \centering
%  \includegraphics[width=2.6in]{grid.pdf}\\
%  \caption{Illustration of the grid partition of true state domain $X$ (the blue shaded area).  The small circles correspond to the $X$-projections of the discrete state from $Q$. For example, the highlighted circle represents the $X$-projection of a discrete state $q$, which is associated with a rectangular region $R_q^x$ in the continuous state space. Region $R_q^x$'s projections onto the one-dimensional space are $R_q^{x_1}$ and $R_q^{x_2}$, respectively, which are marked with dark shaded areas. }\label{fig:grid}
%\end{figure}

Next, we define the transition relation $\tau$ of the raw abstraction $TS^{\rm R}$. 
To this end, we first recursively define the $k$-step reachable set $\textbf{Reach}^k(R_q^x, s)$ of the system's true state $x \in R_q^x$, under control action $s \in \{\text{on}, \text{off}\}$: 
%\begin{align}
%\textbf{Reach}^k(R_q^x, a) = \bigg\{x(k) \bigg\vert \begin{array}{l} x(i+1) = A_a x(i) + K_a, \\ x(0) \in R_q^x\end{array} \bigg\}
%\end{align}
\begin{align}
\textbf{Reach}^1(R_q^x, s) & = \{A_s x + K_s + d \mid x \in R_q^x, d \in D\}, \\
\textbf{Reach}^{k+1}(R_q^x, s) & = \textbf{Reach}^1(\textbf{Reach}^k(R_q^x, s), s).  
\end{align}
Particularly, $\textbf{Reach}^1(R_q^x, s)$ (and hence $\textbf{Reach}^{k+1}(R_q^x, s)$  by induction) can be easily computed as a zonotope given that $R_q^x$ and disturbance set $D$ are zonotopes \cite{girard2005reachability}.  In our case, $R_q^x$ is a rectangle and $D$ is chosen to be a 1-norm ball and hence both are zonotopes. 
 
Given the above definitions for the reachable sets, for any $q, q'\in Q$ and $a\in A = \{1,2\}$, the transition map $\tau$ is defined as follows:
\begin{itemize}
\item For $q \notin Q_{\rm target}$,  $q_{\rm out} \in \tau(q, 1)$ iff 
\begin{align}
& \big((X_{\rm inter}^q \cap X_{\rm target} \neq \emptyset) \wedge (X_{\rm final}^q \not\subseteq X_{\rm target})\big) \nonumber \\
& \ \ \ \vee  (X_{\rm inter}^q \not\subseteq X), 
\label{eq:qout1}
\end{align}
where
\begin{align}
& X_{\rm inter}^q \doteq \Bigg(\bigcup_{k=1}^{N_{\rm dwell}}\textbf{Reach}^k(R_q^x, \text{on})\Bigg) \nonumber \\
& \cup  \Bigg(\bigcup_{k=1}^{N_{\rm grace}}\textbf{Reach}^k\Big(\textbf{Reach}^{N_{\rm dwell}}(R_q^x, \text{on}), \text{off}\Big)\Bigg), \\
& X_{\rm final}^q \doteq \textbf{Reach}^{N_{\rm grace}}\Big(\textbf{Reach}^{N_{\rm dwell}}(R_q^x, \text{on}), \text{off}\Big). 
\end{align}
%Recall that under mode 1, the system first stay in the on-mode for $N_{\rm dwell}$ steps and then stay at the off-mode for $N_{\rm grace}$ steps. Thus the right hand side of Eq. \eqref{eq:qout} is the sequence of intermediate reachable states under mode 1. If not all of these intermediate states fall in domain $X$, we add transition from $q$ to $q_{\rm out}$ to indicate that applying mode 1 at state $q$ is considered to be unsafe. Note that Eq. \ref{eq:qout} is only used for $q\notin Q_{\rm target}$, while $q \in Q_{\rm out}$ needs a separate treatment that will be discussed next.
\item For $q \in Q_{\rm target}$, $q_{\rm out} \in \tau(q, 1)$ iff  $X^q_{\rm inter} \not\subseteq X_{\rm target}$. 
% First note that $X_{\rm target} \subseteq X$. Hence there is a transition from $q$ to $q_{\rm out}$ if some intermediate reachable states go outside the domain $X$, just like the case in the first bullet. Moreover, we add extra transitions from $q$ to $q_{\rm out}$ if some intermediate states only leave $X_{\rm target}$ but not necessarily $X$. This is because we want to stay in the target set once arrived there and hence leaving target is also considered to be unsafe. 
\item $q' \in \tau(q,1)$ for $q'\in Q\setminus \{q_{\rm out}\}$ iff 
\begin{align}
& X_{\rm inter}^q \cap R_{q'}^x \neq \emptyset, \label{eq:q11}\\
& \Big(\textbf{Reach}^{N_{\rm grace}}(R_q^x, \text{on}) \Big)^{x_2}\cap R^{\hat{x}_2}_{q'} \neq \emptyset, \label{eq:q12}\\
& R_q^{\hat{x}_1} = R_{q'}^{\hat{x}_1}.\label{eq:q13}
\end{align}
% The reachable set before ``$\cap$'' in Eq. \eqref{eq:q11} consists of the final reachable states of set $R_q^x$ when the system is about to leave mode 1, and Eq. \eqref{eq:q11} indicates that some of the final reachable states fall in $R_{q'}^{x}$.  Note that we ignore the intermediate reachable states as computed in Eq. \eqref{eq:qout} as long as they are safe (i.e., do not go out $X$ or $X_{\rm target}$ if $q\in Q_{\rm target}$), which is handled by the first two bullets. This practice will not harm the correctness of the obtained controller as reach-stay the target is of our concern. 
% Everything before ``$\cap$'' in Eq. \eqref{eq:q12} is the $x_2$-projection of the reachable states after the system stay in the on-mode for $N_{\rm grace}$, when the measurement of $x_2$ happens. 
% $(\cdot)^{x_2}$ represents the projection onto $x_2$ axis. 
% Eq. \eqref{eq:q12} says that the updated measurement should intersect with $R_{q'}^{x_2}$. 
% Eq. \eqref{eq:q13} says that $\hat{x}_1$ states at $q$ and $q'$ should be the same as $x_1$ is never updated during mode 1. 
\item $q_{\rm out} \in \tau(q, 2)$ iff 
\begin{align}
& \textbf{Reach}^1(R_q^x, \text{off})\not\subseteq X; 
\label{eq:qout2}
\end{align}
% This is essentially the same as the first bullet except that the system can stay in mode 2 for any long time so the reachable set intersection is checked every time step. 
\item $q' \in \tau(q,2)$ for $q'\in Q\setminus \{q_{\rm out}\}$ iff
\begin{align}
& \textbf{Reach}^1(R_q^x, \text{off})  \cap R_{q'}^x \neq \emptyset, \label{eq:q21} \\
& \Big(\textbf{Reach}^1(R_q^x, \text{off}) \Big)^{x_1} \cap R_{q'}^{\hat{x}_1} \neq \emptyset, \label{eq:q22}\\
& R_q^{\hat{x}_2} = R_{q'}^{\hat{x}_2}. \label{eq:q23}
\end{align}
% Eq. \eqref{eq:q21}-\eqref{eq:q23} are essentially the same as Eq. \eqref{eq:q11}-\eqref{eq:q13}, except that i) the reachable set intersection is checked every time step, ii)  $\hat{x}_1$ is updated every time step (Eq. \eqref{eq:q22}), and iii) $\hat{x}_2$ is never updated in mode 2 (Eq. \eqref{eq:q23}). 
\item $\tau(q_{\rm out}, 1) = \tau(q_{\rm out},2) = \{q_{\rm out}\}$
\end{itemize}

\subsubsection{Relation between Raw Abstraction $TS^{\rm R}$ and Hybrid Automaton $\mathcal{H}$} 
Through the above construction, we obtain a finite transition system $TS^{\rm R}$ that ``captures" the behaviors of hybrid automaton $\mathcal{H}$, with some intermediate behaviors being reasonably omitted for reducing spurious behaviors.
Such omission leads to a relation between $\mathcal{H}$ and $TS^{\rm R}$ that is different from the usual behavior over approximation, but will not harm the correctness of the controller obtained through synthesis on the raw abstraction. 
In particular, this relation is specific to solving reach-avoid-stay game with specification $\Phi$ defined in Eq. \eqref{eq:spec}. 
In what follows, we formally state this relation. 
To this end, several definitions are needed.
\begin{defn}
An admissible controller $C: \mathcal{X}_{\rm win }\times \{0,\dots,N_{\rm dwell}+N_{\rm grace}\}$ for hybrid automaton $\mathcal{H}$, where $\mathcal{X}_{\rm win} \subseteq X\times \hat{X}$, is ``induced'' from a controller $\overline{C}:Q_{\rm win} \subseteq Q\rightarrow 2^A$ defined on the raw abstraction $TS^{\rm R}$ if 
$\mathcal{X}_{\rm win} = \bigcup_{q\in Q_{\rm win}}R_q$ 
and $C(x,\hat{x},t_{\rm d}) = \overline{C}(q)$ when $t_{\rm d} = N_{\rm dwell}+N_{\rm grace}$, where $q$ is such that $[x,\hat{x}]^T\in R_q$. 
In particular, $\mathcal{X}_{\rm win}$ and $Q_{\rm win}$ are called ``winning sets'', which only consist good initial conditions starting from where achieving the specification is possible\footnote{For general LTL specifications, the controller needs memory and the winning set may be different from the domain of the controller. However, the reach-stay-avoid specification $\Phi$ in Eq. \eqref{eq:spec} can be achieved by a state feedback controller whose domain is the same as the winning set.}. 
\end{defn}

\begin{defn}\label{defn:vprefix}
Mapping $\Pref: (2^{AP})^{\omega}\rightarrow  (2^{AP})^{\omega} \cup  (2^{AP})^{\ast}$ maps a word $\textbf{w}$ to its ``bad prefix'' for the LTL formula $\Phi$ defined in Eq. \eqref{eq:spec}, i.e., 
\begin{align}
\Pref(\textbf{w}) = 
\begin{cases}
w(1)w(2)\dots w(k^{\ast}_{\textbf{w}}) & \text{ if } k^{\ast}_{\textbf{w}} < \infty\\
\textbf{w} & \text{ otherwise }
\end{cases}.
\end{align}
where 
\vspace{-1.5mm}
\begin{align}
& \ \ \ \ \ \ \ k^{\ast}_{\textbf{w}} \doteq \min \{k\mid \texttt{safe}\notin w(k) \text{, or } \nonumber\\
& \ \ \ \texttt{target}\in w(k-1) \text{ and } \texttt{target}\notin w(k)\}. \label{eq:kstar}
\end{align} 
\end{defn}

To understand Defn. \ref{defn:vprefix}, note that $\Phi$ can be rewritten as 
%$\Phi = \Phi_{\rm safety} \wedge \Phi_{\rm liveness}$ where $ \Phi_{\rm safety} = (\Box \texttt{safe}) \wedge (\neg \texttt{target} \ \mathcal{W} \ \Box \texttt{target})$ and $\Phi_{\rm liveness} = \lozenge \texttt{target}$. 
{\small
\begin{align}
\underbrace{ (\Box \texttt{safe}) \wedge (\neg \texttt{target} \ \mathcal{W} \ \Box \texttt{target})}_{\Phi_{\rm safety}} \wedge \underbrace{(\lozenge \texttt{target})}_{\Phi_{\rm liveness}}. 
\end{align}
}%
Given a word $\textbf{w} = w(1)w(2)\dots \in (2^{AP})^{\omega}$ that violates $\Phi_{\rm safety}$, the violation can be localized to a prefix of $\textbf{w}$ satisfying either $\texttt{safe}\notin w(k)$ or $\texttt{target}\in w(k-1)$ while  $\texttt{target}\notin w(k)$. With this observation, one can talk about ``the first violation'' of $\Phi_{\rm safety}$ that occurs at $k^{\ast}_{\textbf{w}}$ defined in Eq. \eqref{eq:kstar}. 
An easy implication of Defn. 2 is that $\textbf{w}\models \Phi_{\rm safety} \Leftrightarrow \Pref(\ww)$ being infinite.

\begin{defn}
The ``intermediate action projection'' $\textbf{Proj}: A^{\omega}\rightarrow  A^{\omega}$ is a partial mapping defined for any action sequence $\textbf{a} = a(1)a(2)a(3)\dots$ that is admissible to $\mathcal{H}$. 
$\Proj(\aaa)$ replaces any maximal  1-fragment\footnote{See Section \ref{sec:sys_model} for definition. Since $\aaa$ is admissible to $\mathcal{H}$, we know that such maximal 1-fragments of $\aaa$ must consist of $l$ smaller fragments of length $N_1$, where $l$ is arbitrary natural number.} of $\aaa$, e.g. $a(k)a(k+1)\dots  a(k+lN_1-1)$, by the first elements in each $N_1$ sub-fragment, i.e., by $a(k)a(k+N_1)\dots a(k+(l-1)N_1)$. 
Let  $\textbf{w}(\textbf{x}^{\textbf{a},\textbf{d}}) = w(1)w(2)w(3)\dots$ be one word associated admissible $\aaa$.
With a slight abuse of notation, we use $\textbf{Proj}\big(\ww(\xx^{\aaa,\dd})\big)$ to denote the sequence obtained by replacing $w(k)w(k+1)\dots  w(k+lN_1-1)$ (i.e., any fragment associated with a maximal 1-fragments in $\aaa$) by $w(k)w(k+N_1)\dots w(k+(l-1)N_1)$.
%{\small
%\begin{align}
%\underbrace{a(k-1)}_{=2}\underbrace{a(k)a(k+1)\dots  a(k+lN_1-1)}_{=1}
%\label{eq:frag}
%\end{align}
%}%
%by 
%{\small
%\begin{align}
%& \underbrace{a(k-1)}_{=2}\underbrace{a(k)a(k+N_1)\dots a(k+(l-1)N_1)}_{=1}
%\label{eq:frag1}
%\end{align}
%}%
%where $N_1 = N_{\rm dwell} + N_{\rm grace}$, and $\textbf{Proj}(\textbf{w}\big(\textbf{x}^{\textbf{a},\textbf{d}})\big)$ removes any $w(k)$ whose associated $a(k)$ are removed by $\textbf{Proj}(\textbf{a})$.
%In particular, the $a(k-1)=2$ at the very beginning of Eqs. \eqref{eq:frag} and \eqref{eq:frag1}  can be dropped if $k=1$. 
\end{defn}

Observe that $\textbf{Proj}(\textbf{w}) \models \Phi_{\rm liveness} \Rightarrow \textbf{w}\models $$\Phi_{\rm liveness}$.

With the definitions above, the relation between hybrid automaton $\mathcal{H}$ and its raw abstraction $TS^{\rm R}$ can be formally stated by the following proposition. 
\begin{prop}\label{prop:HTSR}
Given any controller $\overline{C}:Q\rightarrow 2^A$ defined on $TS^{\rm R}$, let $C$ be the admissible controller induced by $\overline{C}$. 
For any execution $\textbf{x}^{\textbf{a},\textbf{d}} = \big(x(1),\hat{x}(1)\big)\big(x(2),\hat{x}(2)\big)\big(x(3),\hat{x}(3)\big)\dots$ generated by $\mathcal{H}$ starting from  $t_{\rm d} = N_1$ under controller $C$, 
there exists an execution $\textbf{q}^{\overline{\textbf{a}}} = q(1)q(2)q(3)\dots$, generated by $TS^{\rm R}$  under controller $\overline{C}$, such that
\begin{enumerate}[nolistsep]
\item[(i)] $[x(1),\hat{x}(1)]^T \in R_{q(1)}$, %  for all $k$, 
% \item[(ii)] $\overline{\textbf{a}} =\textbf{Proj}(\textbf{a})$, 
\item[(ii)] $\textbf{Proj}\Big(\Pref\big(\textbf{w}(\textbf{x}^{\textbf{a},\textbf{d}})\big)\Big)  = \Pref\big(\ww(\qq^{\overline{\aaa}})\big)$. 
\end{enumerate}
\end{prop}

By Proposition \ref{prop:HTSR}, we have the following useful result, which guarantees that the controller synthesized for raw abstraction $TS^{\rm R}$ will lead to correct behavior when its induced admissible controller is applied to the hybrid automaton $\mathcal{H}$. 

\begin{thm}\label{thm:HTSR}
Suppose a winning set $Q_{\rm win}\subseteq Q$ and a controller $\overline{C}: Q_{\rm win}\rightarrow 2^{AP}$ is found for $TS^{\rm R}$, so that any execution $\textbf{q}^{\overline{\textbf{a}}} = q(1)q(2)q(3)\dots$ with $q(1)\in Q_{\rm win}$, which is generated under action sequence $\overline{\textbf{a}} = \overline{a}(1)\overline{a}(2)\overline{a}(3)\dots$ satisfying $\overline{a}(k)\in \overline{C}\big(q(k)\big)$, satisfies $\Phi$. 
Then any execution $\textbf{x}^{\textbf{a},\textbf{d}} = \big(x(1),\hat{x}(1)\big)\big(x(2),\hat{x}(2)\big)\big(x(3),\hat{x}(3)\big)\dots$ of $\mathcal{H}$ starting from $[x(1),\hat{x}(1)]^T \in \bigcup_{q\in Q_{\rm win}}R_{q}$ satisfies $\Phi$ under control sequence $\textbf{a} = a(1)a(2)a(3)\dots$ with $a(k)\in C\big(x(k), \hat{x}(k), t_{\rm d}\big)$, where $C$ is induced from $\overline{C}$. 
\end{thm}

%The proofs of Proposition \ref{prop:HTSR} and Theorem \ref{thm:HTSR} can be found in the longer version available at \url{https://bit.ly/2SLV2WU}.
The proofs of Proposition \ref{prop:HTSR} and Theorem \ref{thm:HTSR} are given in the Appendix.

% Constructing the belief space
\subsection{Belief Space Abstraction Construction}
We cannot directly synthesize the controller on the raw abstraction obtained above.
The reason is: since the abstraction (i.e., the finite transition system) is nondeterministic, we cannot locate the actual discrete state on this  raw abstraction due to lack of immediate information of the true state $x$. 
% based on the out-of-date measurement obtained in the event-triggered manner, while operating the system in real-time. 
As a result, should a controller $\overline{C}$ be found on the raw abstraction, we would not be able to follow the action suggested by this controller at the current state, as the precise current state on the raw abstraction is not known.  
Instead, we construct a belief space abstraction, each of whose states may contain multiple raw abstraction states, but can be distinguished from others using the current value of the measured state $\hat{x}$. 
Suppose a controller can be found on the belief space abstraction, we can achieve the control objective with $\hat{x}$.

In what follows we will focus on constructing the belief space abstraction $TS^{\rm B}$ from the raw abstraction $TS^{\rm R}$. 
The belief space abstraction construction relies on a key function $\beta$ that maps a set of raw states to a set of belief states.  
Recall that the set of raw states $Q$ (except for $q_{\rm out}$) can be expressed as a set of tuples $(x^i, \hat{x}^j)$ from the product grid $\{x^1, x^2,\dots, x^N\}\times \{\hat{x}^1, \hat{x}^2,\dots, \hat{x}^N\}$.
Since $x$ is not accessible, the state of the belief space abstraction should be in form of $(S, \hat{x}^j)$, where $S \subseteq \{x^1, x^2,\dots, x^N\}$. 
Formally, let $P$ be the set of states of the belief space abstraction, the state mapping $\beta: 2^{Q} \rightarrow 2^{P}$  is defined as follows: 
\begin{itemize}
\item $\beta(\emptyset) = \emptyset$, $\beta(\{q_{\rm out}\}) = \{p_{\rm out}\}$, 
\item $\beta\Big(\big\{(x, \hat{x})\mid x\in S, \hat{x} = \hat{x}^j \big\}\Big) = \big\{(S, \hat{x}^j)\big\}$, 
\item $\beta(\mathcal{S}) = \bigcup_{\hat{x}^j}\beta\Big(\big\{(x, \hat{x})\in \mathcal{S}\mid  \hat{x} = \hat{x}^{j}\big\}\Big)$.
\end{itemize}
Informally, map $\beta$ collects the raw states indistinguishable from each other with $\hat{x}$ and collapse them into one belief state.   
\begin{wrapfigure}{l}{0\textwidth}
%\centering
  \includegraphics[width=0.2\textwidth]{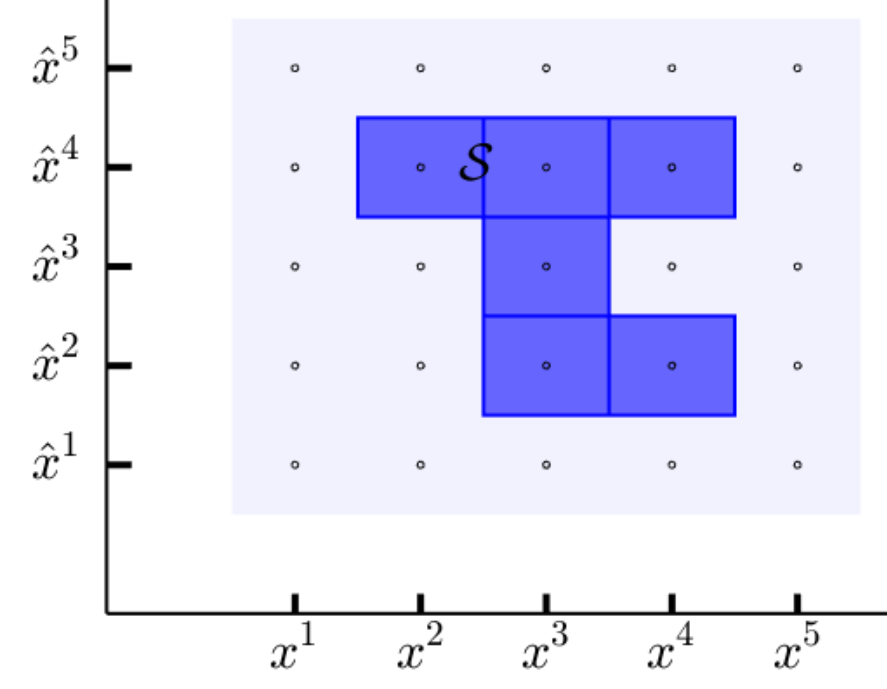}\\
  \caption{The grid partition and a set of raw abstraction states $\mathcal{S} \subseteq Q$. }\label{fig:beta}
  \vspace{-0.3cm}
\end{wrapfigure}
On the other hand, if two raw states are with different $\hat{x}$, they will be collapsed into different belief  states. 
Fig. \ref{fig:beta} shows an example to understand the definition of mapping $\beta$. For ease of illustration, we only use an example where $X$ and $\hat{X}$ are one-dimensional. The heavily shaded region covers a set $\mathcal{S}$ of raw abstraction states. In this example,   
$\beta(\mathcal{S}) = \Big\{\big(\{x^3, x^4\}, \hat{x}^2 \big), \big(\{x^3\}, \hat{x}^3 \big), \big(\{x^2, x^3, x^4\}, \hat{x}^4 \big) \Big\}$. 

% ({\color{red} explanation needed: emphasize that different $x_1$, $\hat{x}_2$ gives different belief space state, visualization may help})

With the state map $\beta$ defined above, we present the belief space abstraction construction with Algorithm \ref{alg:Belief}. 

We start with initial raw state set $Q_{\rm init}$ that corresponds to the continuous state regions containing the given initial continuous state from $X_{\rm init}$. We then map $Q_{\rm init}$ to its corresponding set of belief state $\beta(Q_{\rm init})$ and expand this set with its forward reachable states $P_{ij}$. Note that by definition of $\beta$,  each expanded state from $P_{ij}$ are distinguishable from the rest as they have different values of $\hat{x}$. This procedure will be repeated until no other new states are added in. We know the procedure will terminate because there are only finitely many states in the belief space abstraction. 
%\begin{algorithm}
%\caption{$TS^{\rm B} = \textbf{ConstructBelief}(TS^{\rm R}, \beta, q_{\rm init})$}
%\begin{algorithmic}[1]
%  \Require Raw abstraction $TS^{\rm R} = (Q, A, AP, \lambda, \tau)$, state mapping $\beta: 2^Q \rightarrow 2^P$, initial state $Q_{\rm init}$.
%  \Ensure  Belief space abstraction $TS^{\rm B} = (P, A, AP, \mu, \sigma)$. 
%  \State $P = \emptyset$
%  \State $P' = \beta(Q_{\rm init})$ 
%  \While{$P \neq P'$}
%    \State $P^+ = P'\setminus P$
%    \State $P = P'$
%    \For{$p_i \in P^+$}
%	\For{$a_j \in A$}
%		\State $Q_i = \beta^{-1}\big(\{p_i\}\big)$
%		\State $Q_{ij} = \tau(Q_i, a_j)$
%		\State $P_{ij} = \beta(Q_{ij})$
%		\State $\sigma(p_i, a_j) = P_{ij}$
%		\State $P' = P' \cup P_{ij}$
%	\EndFor
%    \EndFor
%  \EndWhile 
%  \State $P = P\cup \{p_{\rm out}\}$, $\texttt{safe} \in \mu(p)$ iff $p\neq p_{\rm out}$, $\texttt{target}\in \mu(p)$ iff $\texttt{target}\in \lambda(q)$ for all $q\in \beta^{-1}(p)$. 
%  \State \Return $TS^{\rm B} = (P, A, AP, \lambda, \sigma)$
%\end{algorithmic}\label{alg:Belief}
%\end{algorithm}
\begin{algorithm}[b]
\caption{$TS^{\rm B} = \textbf{ConstructBelief}(TS^{\rm R}, \beta, q_{\rm init})$}
\begin{algorithmic}[1]
  \REQUIRE Raw abstraction $TS^{\rm R} = (Q, A, AP, \lambda, \tau)$, state mapping $\beta: 2^Q \rightarrow 2^P$, initial state $Q_{\rm init}$.
  \ENSURE  Belief space abstraction $TS^{\rm B} = (P, A, AP, \mu, \sigma)$. 
  \STATE $P = \emptyset$, \ $P' = \beta(Q_{\rm init})$ 
  \WHILE{$P \neq P'$}
    \STATE $P^+ = P'\setminus P$, \ $P = P'$
    \FOR{$p_i \in P^+$}
	\FOR{$a_j \in A$}
		 \STATE $Q_i = \beta^{-1}\big(\{p_i\}\big)$, \ $Q_{ij} = \tau(Q_i, a_j)$
		 \STATE$P_{ij} = \beta(Q_{ij})$, \ $\sigma(p_i, a_j) = P_{ij}$
		 \STATE $P' = P' \cup P_{ij}$
	\ENDFOR
    \ENDFOR
  \ENDWHILE
   \STATE $P = P\cup \{p_{\rm out}\}$, $\texttt{safe} \in \mu(p)$ iff $p\neq p_{\rm out}$, $\texttt{target}\in \mu(p)$ iff $\texttt{target}\in \lambda(q)$ for all $q\in \beta^{-1}(p)$. 
   \RETURN $TS^{\rm B} = (P, A, AP, \mu, \sigma)$
\end{algorithmic}\label{alg:Belief}
\end{algorithm}

The buck converter control problem is then solved as a reach-stay-avoid game on the belief space abstraction with the same specification $\Phi$. 
Details about the control synthesis algorithm can be found in \cite{nilsson2017augmented}. 
The synthesis algorithm will return 1) a wining set $P_{\rm win} \subset P\setminus \{p_{\rm out}\}$ , and 2) a controller $\overline{\overline{C}}: P_{\rm win} \rightarrow 2^A$  under which $\Phi$ is satisfied for execution starting from $P_{\rm win}$. 
If $\beta(Q_{\rm init}) \subseteq P_{\rm win}$, we claim the control problem of hybrid automaton $\mathcal{H}$ is solved. 
Next we give a definition and formally state this result. 

\begin{defn}
Let $\overline{\overline{C}}:P\rightarrow 2^A$ be a state feedback controller defined for $TS^{\rm B}$, we call a controller $\overline{C}:Q\rightarrow 2^A$ defined for $TS^{\rm R}$ to be ``induced'' from $\overline{\overline{C}}$ if $\overline{C}(q) = \bigcup_{p\in P: q\in \beta^{-1}\big(\{p\}\big)}\overline{\overline{C}}(p)$. 
\end{defn}

\begin{prop}\label{prop:TSRTSB}
Given any controller $\overline{\overline{C}}:P\rightarrow 2^A$ defined on $TS^{\rm R}$, let $\overline{C}$ be the controller induced by $\overline{\overline{C}}$. 
For any execution $\textbf{q}^{\overline{\textbf{a}}} = q(1)q(2)q(3)\dots$ generated by $TS^{\rm R}$ under controller $\overline{C}$, 
there exists an execution $\textbf{p}^{\overline{\overline{\textbf{a}}}} = p(1)p(2)p(3)\dots$, generated by $TS^{\rm B}$  under controller $\overline{\overline{C}}$, such that
\begin{enumerate}[nolistsep]
\item[(i)] $q(1) \in \beta^{-1}\big(\{p(1)\}\big)$, %  for all $k$, 
% \item[(ii)] $\overline{\textbf{a}} =\textbf{Proj}(\textbf{a})$, 
\item[(ii)] $\big\vert\textbf{Pref}\big(\textbf{w}(\textbf{q}^{\overline{\textbf{a}}})\big)\big\vert = \big\vert\textbf{Pref}\big(\textbf{w}(\textbf{p}^{\overline{\overline{\textbf{a}}}})\big)\big\vert$, where $\vert \textbf{w} \vert$ is the length of sequence $\textbf{w}$. 
\item[(iii)] $\forall k: \texttt{target}\notin \lambda\big(q(k)\big) \Rightarrow \forall k: \texttt{target}\notin \mu\big(p(k)\big)$. 
% either both have finite length or both have infinite length. 
\end{enumerate}
\end{prop}
\begin{thm}\label{thm:TSRTSB}
Suppose a winning set $P_{\rm win}\subseteq P$ and a controller $\overline{\overline{C}}: P_{\rm win}\rightarrow 2^{AP}$ is found for $TS^{\rm B}$, so that any execution $\textbf{p}^{\overline{\overline{\textbf{a}}}} = p(1)p(2)p(3)\dots$ with $p(1)\in P_{\rm win}$, which is generated under action sequence 
$\overline{\overline{\textbf{a}}} = \overline{\overline{a}}(1)\overline{\overline{a}}(2)\overline{\overline{a}}(3)\dots$ 
satisfying $\overline{\overline{a}}(k)\in \overline{\overline{C}}\big(p(k-1)\big)$, satisfies $\Phi$. 
Then any execution $\textbf{q}^{\textbf{a}} =q(1)q(2)q(3)\dots$ of $TS^{\rm R}$ started from $q(1)\in Q_{\rm win}\doteq \beta^{-1}(P_{\rm win})$ will satisfy $\Phi$ under control sequence $\overline{\textbf{a}} = \overline{a}(1)\overline{a}(2)\overline{a}(3)\dots$ with $\overline{a}(k)\in \overline{C}\big(q(k-1)\big)$. 
\end{thm} 

The proof of Proposition \ref{thm:TSRTSB} is quite similar to that of Proposition \ref{thm:HTSR} except that 1) we do not need $\textbf{Proj}$ to remove intermediate states during mode 1, and 2) we do not have exact similar prefix in bullet (ii) because $\beta^{-1}\big(\{p\}\big)$ may contain $q$'s labeled with and without $\texttt{target}$, in which case $\texttt{target} \notin \mu(p)$ by Algorithm 1. 
However, bullets (ii) and (iii) still enable a proof of Theorem \ref{thm:TSRTSB}. 

In what follows, several remarks are provided.
%  regarding to the winning set $P_{\rm win}$ and the usage of controller $\overline{\overline{C}}$. 

\begin{figure}
\centering
  \includegraphics[width=2.7in]{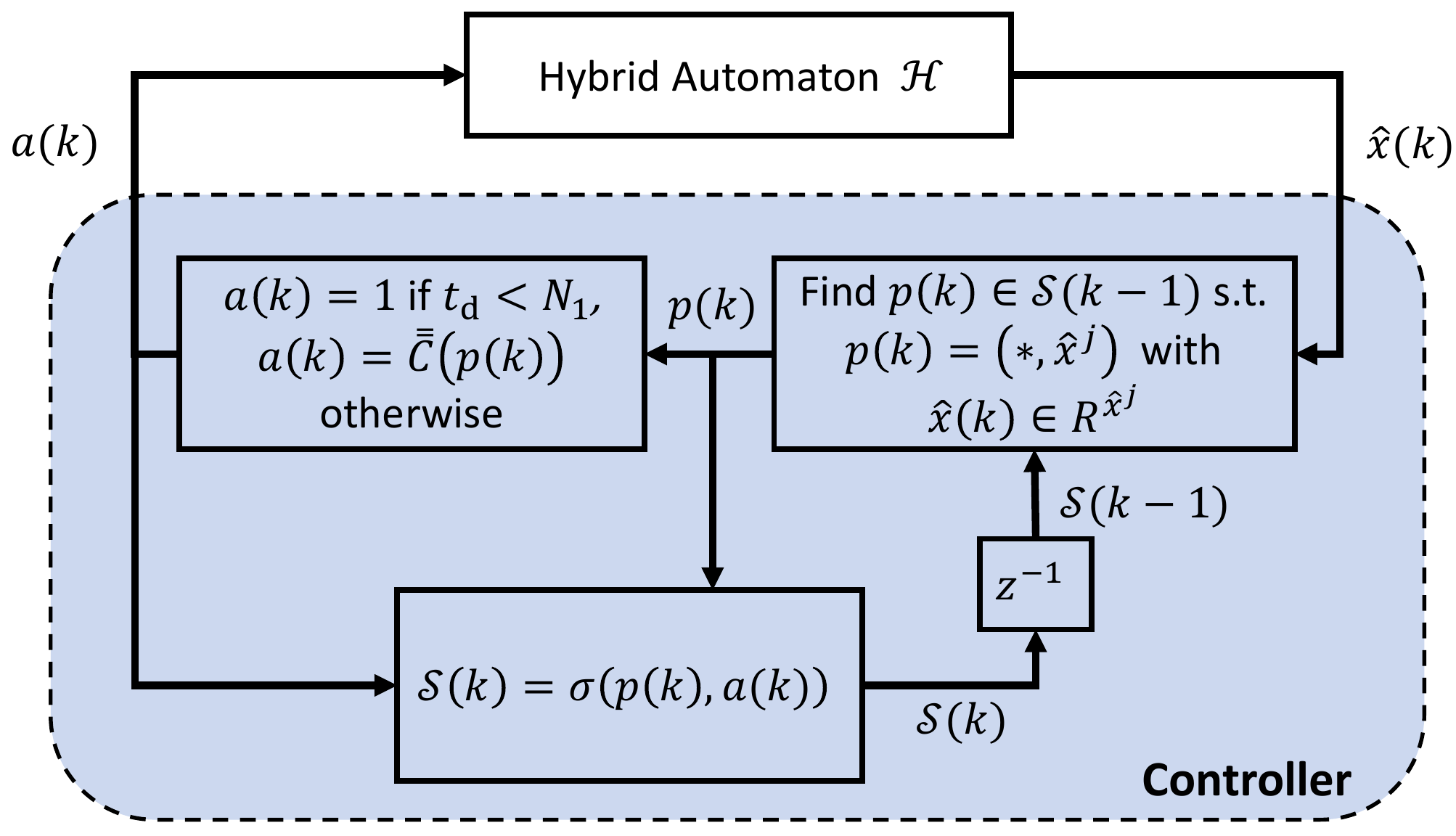}\\
  \caption{Controller Architecture.}\label{fig:TimeLine}
  \vspace{-4mm}
\end{figure}

The first remark is regarding the implementation of the controller $\overline{\overline{C}}$ on the continuous state system in real time.
Although controllers $\overline{C}$ and $C$ can be induced from $\overline{\overline{C}}$, the obtained controller $C$ can not be applied to the hybrid system directly because $C$ still requires both $x$ and $\hat{x}$ to make the decision. 
However, we can use the control structure defined in Fig. \ref{fig:TimeLine}, which only uses the current value of $\hat{x}$.
%It should be noted that the belief space abstraction $TS^{B}$ is not like the regular abstractions (e.g., the raw abstraction in this work), who has a unique state $q$ that corresponds to a state $x$ of the underlying continuous system.   
%Instead, there might be multiple belief states corresponds to continuous state $x$.  Thus we need to decide which belief state $p$ to use to determine the control action $a\in C(p)$. 
Let $p(k)$ be the belief state at time $k$, we set $q(1) = q_{\rm init} \in Q_{\rm init}$ where $p_{\rm init} = \beta\big(\{q_{\rm init}\}\big)$, and $q_{\rm init}$ is such that the given fully measured continuous state $x_{\rm init} \in R_{q_{\rm init}}$.  
We then proceed as follows: 
\begin{enumerate}[nolistsep]
\item[1)] first, we apply action $a(1)$, which can be any element in $\overline{\overline{C}}\big(p(1)\big)$, to the continuous state system;  \item[2)] meanwhile, we need to compute the set  $\sigma\big(p(1),a(1)\big)$ of possible next states on the belief space abstraction; 
\item[3)] since the states in $\sigma\big(p(1),a(1)\big)$ are distinguishable with $\hat{x}$, we can determine the actual next state $p(2)$ using the current value of $\hat{x}$;
%  (possibly already out-of-date, i.e., $\hat{x}\neq x$); 
\item[4)] then we arbitrarily pick control action $a(2)$ from set $\overline{\overline{C}}\big(p(2)\big)$, and  proceed by repeating steps 2) and 3).  
\end{enumerate}
In other words, we need not only the measurement $\hat{x}$, but also $\sigma\big(p(k), a(k)\big)$ to determine control action $a(k+1)$. Hence the controller requires memory. 
The correctness of the control structure presented in Fig. \ref{fig:TimeLine} can be proved by Theorem \ref{thm:HTSR} and \ref{thm:TSRTSB}, noticing that the control actions at any $[x,\hat{x}]^T$ and $t_{\rm d}$ given by this structure is a subset of $C(x,\hat{x},t_{\rm d})$, where $C$ is induced from $\overline{C}$ and $\overline{C}$ is induced from $\overline{\overline{C}}$. 
 
The second remark is about winning set $P_{\rm win}$. Although the belief space abstraction is specific to initial condition $q_{\rm init}$, the obtained controller also works if the system starts from any state $q \in \beta^{-1}(P_{\rm win})$, as a result of the suffix-closedness of $\Phi$ (i.e., $\ww \models \Phi \Rightarrow \ww_k \models \Phi$ for all $k$).
% In this case, the belief space abstraction $TS^{\rm B}(q_{\rm init})$ (the $q_{\rm init}$ in the bracket indicates that the abstraction is associated with initial state $q_{\rm init}$) simulates $TS^{\rm B}(q)$ under each action. This relation holds true because we create the belief state set in a forward expanding way.
% in Algorithm \ref{alg:Belief}.

% ({\color{red}Remarks to be added: 1) using the controller synthesized on the belief space abstraction, why controller need memory; 2) the belief space abstraction is specific to the initial state $q_{\rm init}$})

\vspace{-1mm}
\subsection{Belief Space Size Reduction}\label{sec:reduce} 
%{\color{red}mention exponential size blow up due to subset construction)}

First, in line 10 of Algorithm \ref{alg:Belief}, if $p_{\rm out}\in P_{ij}$, then we can simply set $P_{ij} = \{p_{\rm out}\}$. 
Such replacement is valid because going outside of the continuous state domain is considered unacceptable. 
If $p_{\rm out} \in P_{ij}$, the other states from $P_{ij}\setminus \{p_{\rm out}\}$ will not help solving the synthesis problem but only lead to belief space blow up. 
% ({\color{red}more explanations to be added})

Secondly, note that the reachable set $\textbf{Reach}(R_q^x,a)$ is always a convex set (in fact, it is always a polytope), this suggests that we will not have any belief state $p = (S, \hat{x})$ with a ``disjoint'' $S$. 
Also note that 
% in the buck converter system, $X_2$, the set of noninstantaneously accessible state $x_2$, is just an interval, hence 
$S$ can be overapproximated by $\{x_1^i\mid \underline{i} \leq i\leq \overline{i} \}\times \{x_2^j \mid \underline{j} \leq j\leq \overline{j} \}$. 
Therefore,  instead considering $P \subseteq \big(\{x^1, x^2,\dots, x^N\}\times \{\hat{x}^1, \hat{x}^2,\dots, \hat{x}^N\}\big)\cup\{p_{\rm out}\}$,  we only need to consider  $P \subseteq  \big(\{x^1, x^2,\dots, x^N\}\times \{\hat{x}^1, \hat{x}^2,\dots, \hat{x}^N\}^2\big)\cup\{p_{\rm out}\}$. 
Moreover, $\hat{x}_1$ can be omitted as it only mismatches $x_1$ during mode 1, within which all transitions are autonomous and no control decision is taken. 
Thus the belief space can be further reduced to $P \subseteq \big(\{x^1, x^2,\dots, x^N\}\times \{\hat{x}_2^1, \hat{x}_2^2,\dots, \hat{x}_2^M\}^2\big)\cup\{p_{\rm out}\}$, where $\{\hat{x}_2^1, \hat{x}_2^2,\dots, \hat{x}_2^M\}$ is the grid along $\hat{x}_2$ axis. 
%since 
% ({\color{red}more  explanations to be added})

%% file: sec_Result.tex
%!TEX root = main.tex
We apply the developed approach to the buck converter problems. An alternative design based on a simple open-loop periodic switching rule is used for comparison. 
The desired range of system states are picked as follows: $X = [0, 80] \times [0, 25]$, with target $X_{\rm target} = [0, 80]\times[24, 25]$. 
The specified initial condition is $x_1 = 0, x_2 = 0$. 

% contents
\subsection{Baseline Approach}
We first present a simpler baseline solution approach which only leads to a periodic switching rule that steers the converter voltage into the target range. Since the on-off duration of the switching rule is fixed, this is an open-loop control strategy which does not require any state measurements. 
As will be shown later, however, such open-loop periodic switching leads to undesired transient behavior before the state stabilizes around the target.

Let $N_{\rm on} = N_{\rm dwell}$ be the fixed on-mode duration and $N_{\rm off}$ be the off-mode duration to be determined. Also let $N_{\rm p} = N_{\rm on} + N_{\rm off}$ be the number of steps of one on-off period. The dynamics of the converter's ``periodic behavior'' under such switching rule can be captured by the following higher-dimensional system: 
{\small
\begin{align}
\underbrace{
\left[
\begin{array}{c}
x(1+N_{\rm p}) \\
x(2+N_{\rm p}) \\
\vdots \\
x(2N_{\rm p})
\end{array}
\right]
}_{\xi^+}
= 
\underbrace{
\textbf{diag}(\overline{A}_{k})
}_{\textbf{A}}
\underbrace{
\left[
\begin{array}{c}
x(1) \\
x(2) \\
\vdots \\
x(N_{\rm p})
\end{array}
\right]
}_{\xi}
%\left[
%\begin{array}{c c c c}
%\overline{A}_1 & 0 & \cdots & 0\\
%0  & \overline{A}_2 & \cdots & 0\\
%\vdots   & \vdots & \ddots & \vdots\\
%0 & 0 & \cdots & \overline{A}_{N_{\rm p}} \\
%\end{array}
%\right] 
+ 
\underbrace{
\left[
\begin{array}{c}
\overline{K}_1 \\
\overline{K}_2 \\
\vdots \\
\overline{K}_{N_{\rm p}}
\end{array}
\right]
}_{\textbf{K}}, 
\label{eq:pdym}
\end{align}}%
where $\overline{A}_k = \prod_{i_k = 1}^{N_{\rm p}}A_{i_k}$, $\overline{K}_{k} = \sum_{j_k = 1}^{N_{\rm p}-1}\prod_{i_k = 1}^{N_{\rm p}} A_{i_k}K_{j_k} + K_{N_{\rm p}}$, 
% \begin{align}
% \overline{A}_k = \prod_{i_k = 1}^{N_{\rm p}}A_{i_k}, \ \ \ \overline{K}_{k} = \sum_{j_k = 1}^{N_{\rm p}-1}\prod_{i_k = 1}^{N_{\rm p}} A_{i_k}K_{j_k} + K_{N_{\rm p}}
% \label{eq:}
% \end{align}
and
\begin{align}
A_{i_k} = A_{\rm on}, K_{i_k} = K_{\rm on}, & \text{ if } k + i_k \text{ mod }N_{\rm p} \leq N_{\rm on}, \\ 
A_{i_k} = A_{\rm off}, K_{i_k} = K_{\rm off},  & \text{ otherwise }. 
\end{align}
Since $N_{\rm on} = 10$ is fixed, the objective is to determine $N_{\rm off}$ so that the system in Eq. \eqref{eq:pdym} is asymptotically stable around its equilibrium $\xi_{\rm eq}$ (i.e., $(I - \textbf{A})\xi_{\rm eq} = \textbf{K}$), and all the $x$ fragment in $\xi_{\rm eq}$ fall in the corresponding target range. To achieve this, we do a line search on $N_{\rm off}$ and pick $N_{\rm off} = 35$. 

\begin{figure}[h]
  \centering
  \includegraphics[width=2.95in]{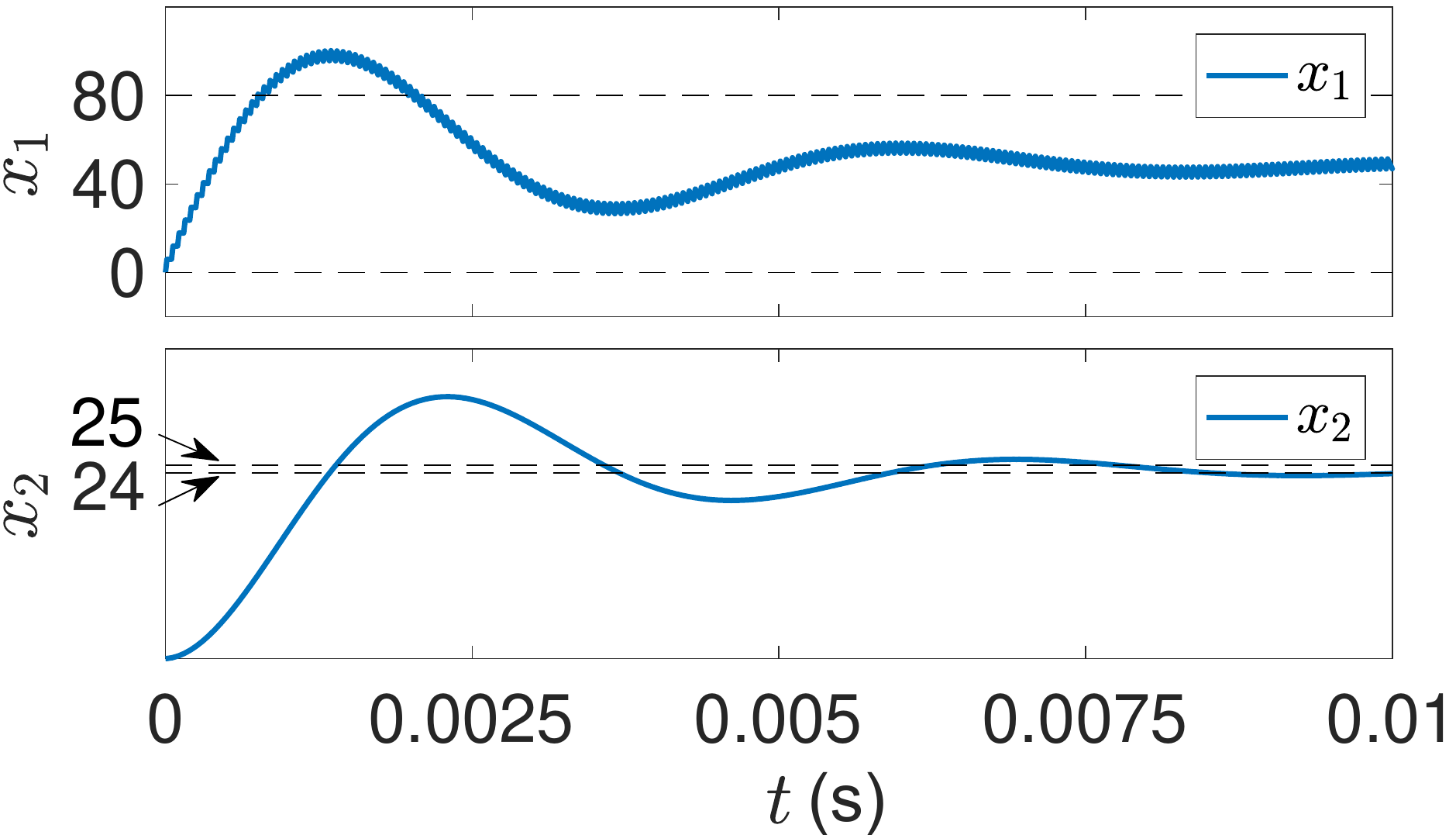}\\
  \caption{Simulation result with the open-loop periodic switching rule. The red dashed lines mark the target range of state $x_1$ and $x_2$. The ``band-like'' $x_1$-plot is due to high-frequency switching.}\label{fig:Sim0}
\end{figure}

Fig. \ref{fig:Sim0} shows the resulting behavior under the open-loop periodic switching rule. It can be seen that the true state $x$ eventually converges into the desired range $X_{\rm target}$, but it experiences a large overshot before stabilizing around the periodic steady state behavior.  
Such large overshot is undesirable and can risk safety. 
This also suggests the necessity of specifying the desired system behavior with temporal logic (e.g., staying within the target once arrived, never going into unsafe region, etc), rather than just requiring stability. 

\subsection{Proposed Approach}
\begin{figure}[h]
  \centering
  \includegraphics[width=3.1in]{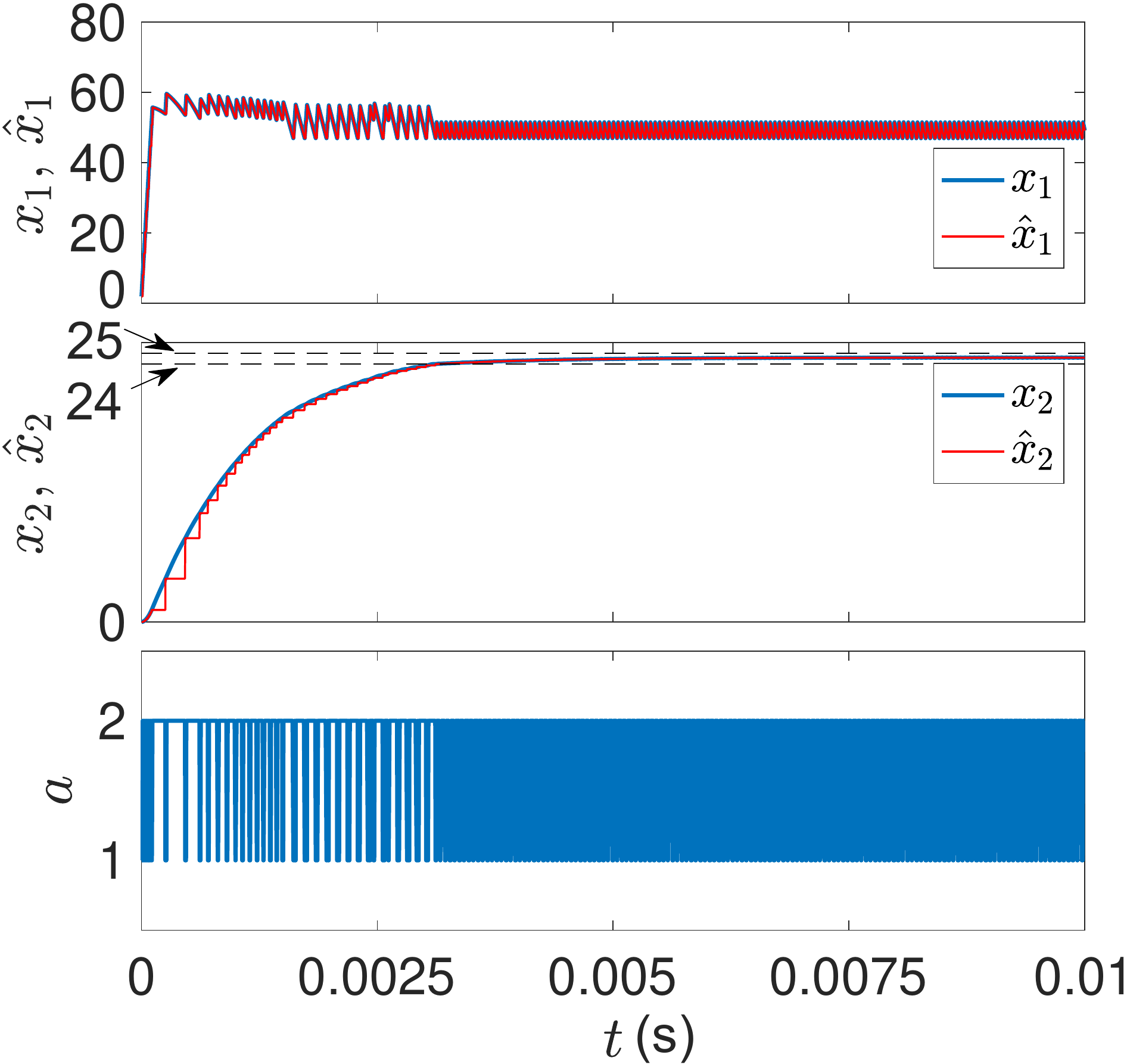}\\
  \caption{Simulation result with the obtained correct-by-construction switching controller. }\label{fig:Sim}
\end{figure}

We apply the proposed solution approach to the buck converter control problem and obtained a controller. 
Fig. \ref{fig:Sim} shows the simulation result of the closed loop system. 
The top plot shows the current $x_1$ and its measurement $\hat{x}_1$. The two variables have very close values and can be hardly distinguished from the plot. The middle plot shows the voltage $x_2$ and its measurement $\hat{x}_2$. It can be seen that $x_2$ converges to the target interval in finite time and stays within the interval once there. The bottom plot shows the switching signal. The update of $\hat{x}_2$ is consistent with the switching signal. We also observe that the switching sequence has a periodic behavior at the steady state (i.e., after settling down in the target interval). This is the result of fixed on-mode dwell time, and some control action selection heuristic, i.e.,  we select the control actions that are more likely to maintain the previous on-off switching pattern whenever there are multiple actions suggested by the controller. 

Next we report the result of belief space size reduction. Without any reduction, the total number of belief space abstraction contains $3.43\times10^{15}$ states. After omitting the measurement $\hat{x}_1$, as explained in Section \ref{sec:reduce}, the number of states reduces to $3.19\times 10^6$. The number is further reduced to $6.59\times 10^4$ after replacing the arbitrary subset construction by the disjoint subset construction (see Section \ref{sec:reduce}). 
% {(\color{blue}mention raw/delief space abstraction size) YES, please add this.}

%% file: sec_Appendix.tex
%!TEX root = main.tex

\subsection{Proof of Proposition \ref{prop:HTSR}}
\begin{proof}
Let $C, \overline{C}, \textbf{x}^{\textbf{a},\textbf{d}}$ be defined as in Proposition \ref{prop:HTSR}, we now construct $\overline{\textbf{a}}$ and $\textbf{q}^{\overline{\textbf{a}}}$ that satisfy bullets (i) and (ii). To this point, we set $q(1)$ to be such that $[x(1),\hat{x}(1)]^T \in R_{q(1)}$. With $t_{\rm d} = N_1$, this implies $C\big(x(1),\hat{x}(1),t_{\rm d}\big) = \overline{C}\big(q(1)\big)$.
Then we evolve $TS^{\rm R}$ under $\overline{C}$ starting from $q(1)$, and show that this can lead to $\overline{\textbf{a}}$ and $\textbf{q}^{\overline{\textbf{a}}}$ that satisfy bullet (ii) by induction. 
As the induction step, we assume that $[x(k),\hat{x}(k)]^T \in R_{q(k')}$, $t_{\rm d} = N_1$ and $C\big(x(k),\hat{x}(k),t_{\rm d}\big) = \overline{C}\big(q(k')\big)$.  
If $a = 2\in C\big(x(k),\hat{x}(k),t_{\rm d}\big)$ and $a(k) = 2$ is used, by Eq. \eqref{eq:q21}-\eqref{eq:q23}, we know that there exists $q(k'+1) \in \tau\big(q(k'),2\big)$ s.t. $[x(k+1),\hat{x}(k+1)]^T\in R_{q(k'+1)}$ and $t_{\rm d}$ remains to be $N_1$. 
If $a = 1\in C\big(x(k),\hat{x}(k),t_{\rm d}\big)$ and $a(k) = 1$ is used, by Eq. \eqref{eq:q11}-\eqref{eq:q13}, there exists $q(k'+1) \in \tau\big(q(k'),2\big)$ s.t. $[x(k+N_1),\hat{x}(k+N_1)]^T\in R_{q(k'+1)}$ and $\lambda\big(q(k'+1)\big) = \ell\big(x(k+N_1)\big)$ due to the proposition preserving property of the partition.
During time steps from $k$ to $k+N_1$, $t_{\rm d}$ will be reset to zero and recount to $N_1$.
Since $C$ is admissible to $\mathcal{H}$, we know that $a(k+1), \dots, a(k+N-1) = 1$ will be removed by $\textbf{Proj}$.  
Hence $\ell\big(x(k+1)) \dots \ell\big(x(k+N-1)\big)$ will also be removed, and we have $\textbf{Proj}\big(\textbf{w}(\textbf{x}^{\textbf{a},\textbf{d}})\big) = \textbf{w}(\textbf{q}^{\overline{\textbf{a}}})$. 
In the case that $\textbf{w}(\textbf{x}^{\textbf{a},\textbf{d}})\models \Phi_{\rm safety}$, this concludes the induction step because in this case, we have $q(k) \neq q_{\rm out}$ for all $k$ and $\Pref$ is just identity mapping for both $\textbf{w}(\textbf{x}^{\textbf{a},\textbf{d}})$ and $\textbf{w}(\textbf{q}^{\overline{\textbf{a}}})$. 
In the case that $\textbf{w}(\textbf{x}^{\textbf{a},\textbf{d}})\not\models \Phi_{\rm safety}$, the above induction can last until the first violation of $\textbf{w}(\textbf{x}^{\textbf{a},\textbf{d}})\not\models \Phi_{\rm safety}$ (see Definition 2) that occurs at $k = k^{\ast}$. In this case $q((k^{\ast})') = q_{\rm out}$ by Eq. \eqref{eq:qout1} and \eqref{eq:qout2}, where $(k^{\ast})'$ is defined similar to  that in the above induction. 
Recall that $\texttt{safe}\notin \lambda(q_{\rm out})$, $\textbf{Pref}\big(\textbf{w}(\textbf{q}^{\overline{\textbf{a}}})\big)$ is the prefix of $\textbf{w}(\textbf{q}^{\overline{\textbf{a}}})$ until  $(k^{\ast})'-1$ position. 
Hence $\textbf{Proj}\Big(\Pref\big(\textbf{w}(\xx^{\aaa,\dd})\big)\Big)  = \Pref\big(\ww(\qq^{\overline{\aaa}})\big)$. 
\end{proof}

\subsection{Proof of Theorem \ref{thm:HTSR}}
\begin{proof}
Let $C, \overline{C}, Q_{\rm win}, \aaa, \xx^{\aaa,\dd}$ be defined as in Theorem \ref{thm:HTSR}. 
By Proposition \ref{prop:HTSR}, there exists $\overline{\aaa}$ and $\qq^{\overline{\aaa}}$ satisfy the two bullets. 
By bullet (i), we know that $q(1)\in Q_{\rm win}$.
% Combining bullet (i), (ii), Definition 1 and the fact that $a(k)\in C\big(x(k-1), \hat{x}(k-1), t_{\rm d}\big)$, we know $\overline{\textbf{a}}$ is such that $\overline{a}(k')\in \overline{C}\big(q(k'-1)\big)$ and hence can be viewed as a  action sequence generated under controller $\overline{C}$. 
Moreover, by definition $\overline{\aaa}$ and  $\qq^{\overline{\aaa}}$ is such that $\overline{a}(k)\in \overline{C}\big(q(k)\big)$.
This implies that $\ww(\qq^{\overline{\aaa}})\models \Phi$, that is, $\ww(\qq^{\overline{\aaa}})\models \Phi_{\rm safety}$ and $\ww(\qq^{\overline{\aaa}})\models \Phi_{\rm liveness}$. 
\begin{enumerate}[nolistsep]
\item[1)] First, $\ww(\qq^{\overline{\aaa}})\models \Phi_{\rm safety}$ means that $\Pref\big(\ww(\qq^{\overline{\aaa}})\big) = \ww(\qq^{\overline{\aaa}})$ is an infinite sequence.
Then we know by bullet (ii) that $\Pref\big(\ww(\xx^{\aaa,\dd})\big)$ must also be infinite and hence $\ww(\xx^{\aaa,\dd})\models \Phi_{\rm safety}$. This also gives $\Pref\big(\ww(\xx^{\aaa,\dd})\big) = \ww(\xx^{\aaa,\dd})$. 
\item[2)] With $\ww(\qq^{\overline{\aaa}})\models \Phi_{\rm liveness}$ and $\Pref\big(\ww(\qq^{\overline{\aaa}})\big) = \ww(\qq^{\overline{\aaa}})$ from 1),  we have $\Pref\big(\ww(\qq^{\overline{\aaa}})\big) \models \Phi_{\rm liveness}  \Rightarrow  \textbf{Proj}\Big(\Pref\big(\ww(\xx^{\aaa,\dd})\big)\Big) \models \Phi_{\rm liveness}$ by bullet (ii). This further implies $\ww(\xx^{\aaa,\dd}) \models \Phi_{\rm liveness}$ because $\Pref\big(\ww(\xx^{\aaa,\dd})\big) = \ww(\xx^{\aaa,\dd})$ and $\textbf{Proj}(\ww)\models \Phi_{\rm liveness} \Rightarrow \ww\models \Phi_{\rm liveness}$. 
\end{enumerate}
Finally we have $\ww(\xx^{\aaa,\dd}) \models \Phi_{\rm safety}\wedge \Phi_{\rm liveness} = \Phi$. 
\end{proof}